\documentclass[aps,onecolumn,11pt,tightenlines,showkeys,superscriptaddress,nofootinbib]{revtex4-2}
\usepackage[T1]{fontenc}
\usepackage{lmodern}
\usepackage{amsmath,amssymb,graphicx}
\usepackage{url}
\usepackage[hidelinks]{hyperref}
\AtBeginDocument{}
\begin{document}
\title{Xenon energy-deposition widths in silicon: measurements and Geant4 comparisons}
\author{Andrei Azhakin}
\email{azhakin.am21@physics.msu.ru}
\affiliation{Faculty of Physics, Lomonosov Moscow State University,
Moscow, 119991 Russia}
\author{Ilya Kudryashov}
\author{Dmitry Karmanov}
\author{Igor Kovalev}
\affiliation{Skobeltsyn Institute of Nuclear Physics,
Lomonosov Moscow State University, Moscow, 119991 Russia}
\begin{abstract}
Energy-deposition fluctuations determine the isotope-separation capability
of multilayer silicon detectors. We measure the central peak widths of
$^{124}\mathrm{Xe}$ signals in 300- and 675-micrometre silicon layers at
incident energies of approximately 152--274 MeV per nucleon and compare
them with Geant4 11.4.0 electromagnetic option 4 (emopt4) and a calculation
with explicit charge exchange. In the 300-micrometre detector, the emopt4
relative width of $1.844\%$
exceeds the unselected experimental width of $1.407\%$ by $31\%$ and the
selected width of $1.060\%$ by $74\%$. For the 675-micrometre detector,
selection windows around the downstream-signal peak, each approximately
ten fitted first-plane standard deviations wide, give experimental upper estimates
of $0.66$--$0.74\%$, compared with emopt4 predictions of
$1.01$--$1.24\%$. Explicit charge-state transport gives widths of
$0.889$--$0.932\%$ for the thinner layer, depending on the incident charge
distribution, and $0.59$--$0.63\%$ for the thicker layer. These predictions
are consistent with the experimental upper estimates, while emopt4
overestimates the widths for both thicknesses throughout the studied
energy interval.
\end{abstract}
\keywords{heavy ions; silicon detectors; energy-deposition fluctuations; Geant4; isotope identification}
\maketitle

\section{Introduction}
The isotope composition of cosmic rays provides information on their
sources and propagation. In a multilayer silicon detector, neighbouring
isotopes are separated through differences in their energy-deposition
profiles. The fluctuations of these signals determine the attainable
mass resolution and the detector thickness needed to reach it.

For partially stripped heavy ions, fluctuations of the ionic charge add
to the collisional straggling at fixed charge. Measurements by Weick
\textit{et al.} with Au, Pb, Bi and U showed that this charge-exchange
contribution can dominate the energy-loss straggling~\cite{Weick2000,Weick2002}.
Its magnitude depends on the capture and loss cross sections and on the
stopping powers of the individual charge states.

Silicon-detector measurements have tested straggling descriptions for
light and medium-mass ions over a broad range of fractional energy
loss~\cite{Hasebe1994,Hasebe1999}. Experimental coverage of deposited-energy
widths remains limited for partially stripped high-$Z$ ions at hundreds
of MeV per nucleon. In a finite detector, secondary-electron transport
and escape also affect the energy retained in the sensitive volume.
Dedicated measurements of signal widths are therefore needed to test
how charge-exchange straggling contributes to detector response.

Geant4~\cite{Agostinelli2003} is widely used to calculate detector response.
We compare xenon measurements from two silicon-detector systems at the
Joint Institute for Nuclear Research in Dubna with standard
electromagnetic option 4 (emopt4) and our Geant4
extension with explicit electron capture and loss. The comparison uses
relative central peak widths. In STIKL, the downstream signals provide
an event selection that reduces broadening from the incident-energy
spread.

\section{Experimental arrangements}
\subsection{DPS-NICA measurements}
\begin{figure}[!htb]
\centering
\includegraphics[width=\linewidth]{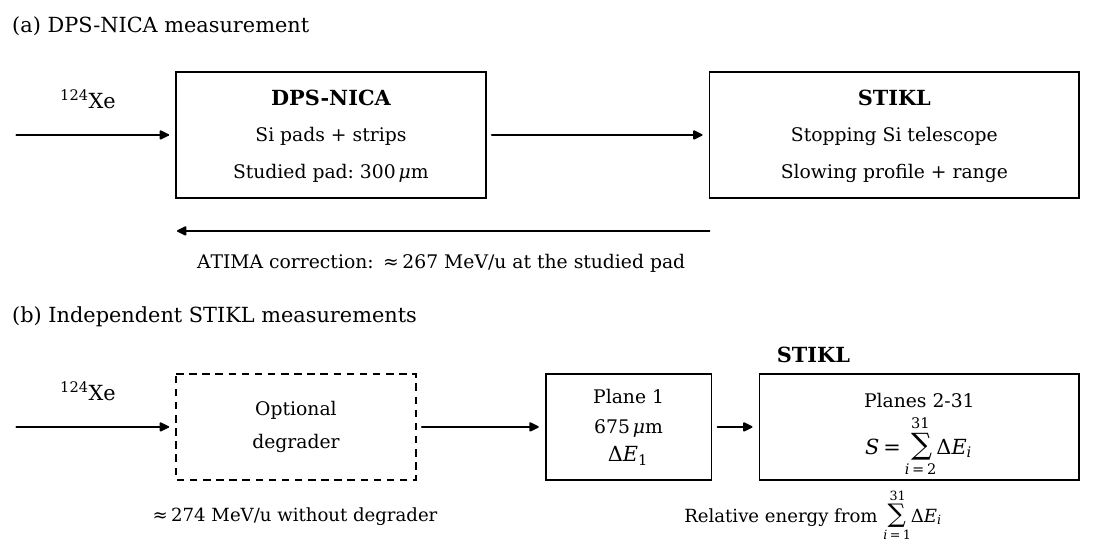}
\caption{Experimental arrangements: (a) DPS-NICA followed by STIKL;
the stopping profile and range in STIKL, together with the intervening
energy losses, determine the energy at the studied DPS-NICA pad;
(b) independent STIKL measurements with and without degraders.
The first-plane signal is analysed, while the sum of planes 2--31 defines
the event selection. The diagrams are not to scale.}
\label{fig:beam_energy}
\end{figure}

The DPS-NICA apparatus comprises silicon pad and strip
detectors~\cite{Kurganov2024}. In these measurements the STIKL silicon
stopping telescope developed for NUCLEON-2~\cite{Vasiliev2021} was placed
downstream [Fig.~\ref{fig:beam_energy}(a)]. An almost monoisotopic
$^{124}\mathrm{Xe}$ beam traversed the detectors close to normal incidence.
The analysed DPS-NICA pad was $300\,\mu\mathrm{m}$ thick.

The absolute energy scale was obtained from ATIMA stopping-power
tables~\cite{ATIMA,ScheidenbergerGeissel1998} using the measured slowing-down
profile and the ion range in silicon. The reconstruction included energy
losses in air and the individual mean thickness of each detector plane.
Correcting for losses between the analysed DPS-NICA pad and the first
STIKL plane gave an energy of approximately $267\,\mathrm{MeV}$ per
nucleon at the pad. For measurements without a degrader, the STIKL
total-energy peak bounds the relative beam-energy spread at less than
$1.5\%$. Electronic noise, evaluated from pedestal distributions, was
small compared with the measured first-plane signal widths.

The unselected DPS-NICA sample contains all $14\,025\,907$ recorded events,
including edge events and nuclear interactions. The selected sample
contains $22\,088$ events with reconstructed tracks through a central
$3\times3\,\mathrm{mm}^{2}$ region and xenon-signal gates in four downstream
strip detectors. These broad gates identify xenon events; their widths
are much larger than the signal fluctuations in the studied pad.
The central region suppresses edge effects and
broadening from thickness variations across the silicon plate.
No amplitude cut was applied to the analysed pad.
Angular maps revealed channeling, and events near the channeling
directions of this pad were excluded. In the first DPS-NICA pad, the
crystallographic axis was nearly aligned with the beam, preventing this
rejection. Channeling there can alter the charge distribution entering
the analysed pad.

\subsection{Independent STIKL measurements}
The STIKL measurements were taken in separate runs without DPS-NICA
upstream, with and without degraders [Fig.~\ref{fig:beam_energy}(b)].
We analyse the first plane, $675\,\mu\mathrm{m}$ thick. The incident energy
without a degrader was approximately $274\,\mathrm{MeV}$ per nucleon,
with the absolute scale determined by the same ATIMA-based procedure.
Channeling was also observed in STIKL, and no angular rejection was applied.
No position measurement was available to reject edge events directly.
Before reaching STIKL, the beam traversed several metres of air and
several thin extraction and beam-pipe windows, with a combined thickness
of order a few hundred micrometres, as well as the degraders when present.
The comparison uses three samples: degraded-beam data at approximately
152 and $195\,\mathrm{MeV}$ per nucleon, and a run without a degrader
recorded on 19 March.

Events are selected using $S$, the sum of signals in planes 2--31,
without a direct cut on the first-plane signal $E_1$. The window centre
$S_c$ is fixed at the xenon maximum of the summed-signal distribution;
the higher-signal maximum is used for the run without a degrader.
The selection restricts the incident-energy range. The window width is determined
iteratively as described in Sec.~\ref{sec:stikl_selection}.

The relative energy scale follows the mean original total signal in
planes 1--31 after selection, normalised to approximately $274\,\mathrm{MeV}$ per
nucleon for the run without a degrader. The selected samples then have
energies of approximately 152, 195 and $274\,\mathrm{MeV}$ per nucleon.

For both data sets, we assign a conservative $\pm5\%$ systematic
uncertainty to the reconstructed energy scale. The published ATIMA 1.4
benchmark for Xe at 60--290 MeV per nucleon gives a weighted relative
standard deviation of $1.5\%$ from the measured stopping powers~\cite{WeickATIMA2017}.
As a sensitivity check, varying the stopping-power scale by $\pm5\%$
and inverting the Xe--Si range tables changes the reconstructed energies
by approximately $\pm3.2\%$ over the studied interval. The adopted
$\pm5\%$ energy uncertainty leaves a margin for the low-energy part of
the stopping profile and the air-loss correction.

\subsection{Energy spread in the STIKL beam}
\label{sec:stikl_beam_spread}
\begin{figure}[!htb]
\centering
\includegraphics[width=\linewidth]{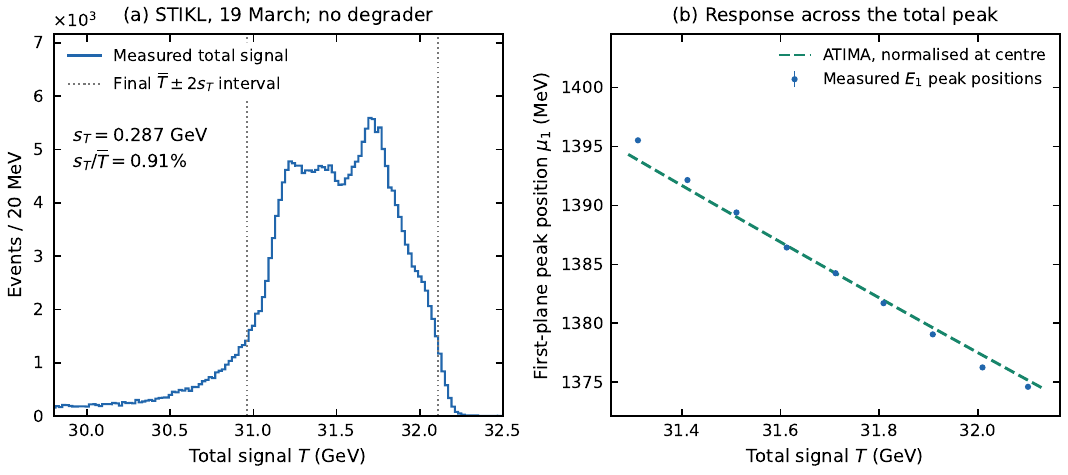}
\caption{Energy-spread diagnostic for the STIKL run without a degrader
at approximately $274\,\mathrm{MeV}$ per nucleon. (a) Original total
signal $T$ in planes 1--31. Dotted lines bound the final
$\overline T\pm2s_T$ interval used to calculate the iterative RMS;
both maxima are retained.
(b) Fitted first-plane peak positions in nonoverlapping
$100\,\mathrm{MeV}$ intervals of $T$, with statistical fit errors.
The ATIMA mean-loss curve is normalised to the central bin.}
\label{fig:stikl_beam}
\end{figure}

The original total signal $T=E_1+S$, before selection by the downstream
sum, provides an estimate of the incident-energy distribution.
We characterise the central width of these non-Gaussian spectra by
an iterative RMS, $s_T$. Starting with a broad $\pm2.2\,\mathrm{GeV}$
interval around the xenon feature, the sample mean $\overline T$ and
RMS are recomputed within $|T-\overline T|<2s_T$ until stable.
The resulting relative widths $s_T/\overline T$ are
$2.13\%$, $1.95\%$ and $0.91\%$ for the samples near 152, 195 and
$274\,\mathrm{MeV}$ per nucleon, respectively.
Figure~\ref{fig:stikl_beam}(a) shows the run without a degrader.

To test the energy interpretation of this spread, we fit the first-plane
peak in nonoverlapping $100\,\mathrm{MeV}$ intervals of $T$.
Its position decreases with increasing total signal, following the
energy dependence of the ATIMA mean losses
[Fig.~\ref{fig:stikl_beam}(b)]. Near the reference energies, the measured
logarithmic slopes $-d\ln\mu_1/d\ln T$ are approximately 0.65, 0.58 and
0.59, compared with 0.66, 0.61 and 0.54 from ATIMA. The normalisation
is matched at the central bin.
Therefore, the central total-signal widths indicate characteristic beam-energy
spreads of about $0.3$--$0.5\,\mathrm{GeV}$ per ion. The corresponding
summed-signal widths are 26--39 times the first-plane widths obtained
from the selected samples (Table~\ref{tab:stikl}).

\newpage
\section{Simulations and width extraction}
\subsection{Calculation settings}
The reference calculations use Geant4 11.4.0 with emopt4 and a single
$32\times32\,\mathrm{mm}^{2}$ silicon plate of density
$2.33\,\mathrm{g\,cm}^{-3}$ in vacuum. A monoenergetic
$^{124}\mathrm{Xe}$ beam is generated $9\,\mathrm{mm}$ before the plate
and directed normally at its centre. The DPS-NICA calculation uses
$267\,\mathrm{MeV}$ per nucleon and $300\,\mu\mathrm{m}$ of silicon;
the STIKL calculations use 154, 197.5 and $267.8\,\mathrm{MeV}$ per
nucleon and $675\,\mu\mathrm{m}$ of silicon. Each sample contains
$10^4$ events.

The reference retains the standard multiple-scattering, nuclear-stopping
and ionisation processes, including the native ion-fluctuation model.
The charge-exchange contribution is described through the empirical
parameterisation based on Yang \textit{et al.}~\cite{Yang1991}.
Nuclear reactions are disabled. Electron transport in silicon uses the
MicroElec elastic and inelastic models~\cite{Valentin2012,Raine2014},
with standard electromagnetic models above $9\,\mathrm{MeV}$.
The production cut is $25\,\mu\mathrm{m}$, corresponding to an
electron-production threshold of approximately $54\,\mathrm{keV}$.
The MicroElec electron-termination threshold is separately set to the
same energy with \texttt{SetKillBelowThreshold}; below it, the track is
stopped and the remaining kinetic energy is deposited locally.
Checks with cuts of 10 and $25\,\mu\mathrm{m}$ changed the peak width by
no more than its statistical uncertainty.

\subsection{Calculation with explicit charge exchange}
The alternative calculation uses our Geant4 extension with explicit
charge exchange (CC). Individual electron-capture and loss events are
sampled from GLOBAL cross sections~\cite{Scheidenberger1998}. The stopping
power follows the instantaneous ionic charge through a $q^2$ dependence,
normalised to the ATIMA equilibrium mean. Charge-exchange fluctuations
then arise from the sampled charge histories, replacing the empirical
contribution in the reference model.

Explicit secondary-electron production includes a Mott correction
using the polynomial approximation of Ref.~\cite{Lijian1995} with the
coefficients supplied in Geant4. Electron transport and the geometry are the same
as in the reference calculation. For xenon near $280\,\mathrm{MeV}$ per
nucleon, switching off this correction produced no observable change in the
deposited-energy peak width. The software implementation and validation
against the measurements of Weick \textit{et al.} are presented in a
companion paper~\cite{CompanionPackage}.

\subsection{Width extraction}
We define the relative central peak width as
$w=\sigma_{\mathrm{core}}/\mu_{\mathrm{core}}$.
All experimental and simulated first-plane peaks are fitted with the
same bin-integrated Poisson likelihood. The expected count in each bin
is obtained by integrating the Gaussian over the bin boundaries, with
a single free normalisation shared by all bins. Empty bins are included.
The fitting interval is adjusted iteratively
to $\mu\pm2\sigma$.
Experimental spectra retain their native ADC channels. For STIKL,
the channel boundaries are converted to energy using the detector
calibration after subtracting the most recently measured pedestal.
Events between successive pedestal measurements are fitted jointly
with common $\mu$ and $\sigma$; each such group has one free normalisation.
Simulated spectra use $1\,\mathrm{MeV}$ bins.

\subsection{STIKL selection window}
\label{sec:stikl_selection}
The gate on the downstream sum must restrict the incident-energy range
while retaining enough events to fit the first-plane peak. It should
also be wide enough to avoid substantial narrowing of this peak even
under ideal energy balance. Therefore, we set its full width to approximately ten
first-plane standard deviations. At fixed $S_c$, the iteration is
\begin{equation}
 W_{k+1}=10\sigma_{1,k},
 \qquad |S-S_c|<W_k/2,
 \label{eq:stikl_gate}
\end{equation}
where $\sigma_{1,k}$ is fitted to the first-plane peak of the selected
events. The iteration stops when the relative change in $W$ is below
$0.5\%$. The final windows retain between $1.1\times10^4$ and
$2.7\times10^4$ events.

We estimate the broadening from the accepted energy range using the ATIMA mean-loss
curve and the total-signal energy calibration. For a locally uniform
incident-energy distribution and a linear mean response,
\begin{equation*}
 \sigma_{\rm mix}\simeq
 \left|\frac{d\overline E_1}{d\overline S}\right|
 \frac{W}{\sqrt{12}}.
\end{equation*}
Here the derivative describes the change of the mean signals with
incident energy. At the adopted windows, $\sigma_{\rm mix}$ is approximately
2.9, 1.5 and $0.7\,\mathrm{MeV}$ for the samples near 152, 195 and
$274\,\mathrm{MeV}$ per nucleon. For the measured first-plane widths,
this corresponds to an estimated broadening of about
$2\%$, $0.8\%$ and $0.2\%$, respectively. Using the measured sum
distribution instead of a uniform density changes $\sigma_{\rm mix}$
by less than $1\%$.

The downstream sum also carries information about first-plane
fluctuations through energy conservation. We estimate the resulting
narrowing in the ideal limit of perfectly measured energy balance,
using a locally Gaussian first-plane response, the ATIMA mean-loss
slope and a broad incident-energy distribution. The latter approximation
is supported by the beam-energy spreads inferred in
Sec.~\ref{sec:stikl_beam_spread}, which exceed the measured first-plane
widths by a factor of several tens. Independent fluctuations of the summed
detector response weaken the energy-balance constraint, making the
ideal-balance estimate of narrowing conservative. For the adopted
$10\sigma_1$ windows, the maximum narrowing in this model is about
$2$--$5\%$: $5.1\%$, $3.7\%$ and $2.1\%$ near 152, 195 and
$274\,\mathrm{MeV}$ per nucleon, respectively. We include this effect
as a one-sided systematic uncertainty towards larger widths,
$\delta w_{\rm sel}^{+}=wf/(1-f)$, where $f$ is the fractional
narrowing. The fitted values are unchanged. With this allowance,
we use the selected widths as upper estimates that retain residual
beam and detector broadening.

\section{Results and model comparison}
\subsection{DPS-NICA spectra}
\begin{figure}[!htb]
\centering
\includegraphics[width=\linewidth]{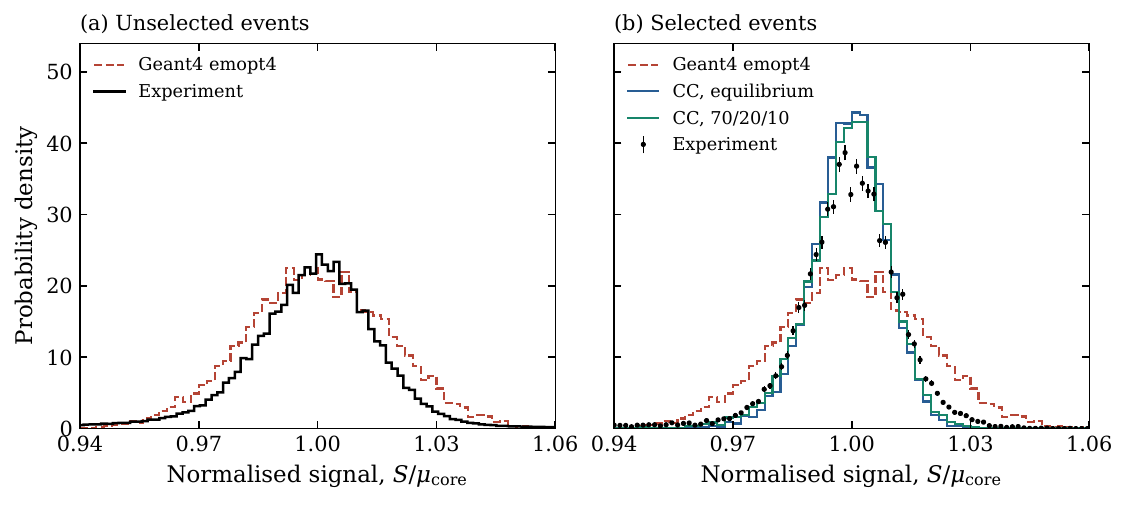}
\caption{DPS-NICA xenon signal distributions at an incident energy of
approximately $267\,\mathrm{MeV}$ per nucleon: (a) the unselected sample
and emopt4; (b) the selected sample, emopt4 and explicit charge-exchange
calculations. The CC samples use either the equilibrium charge distribution
in silicon or incident fractions of $70\%$, $20\%$ and $10\%$ for
$q=54,53,52$. Each signal is divided by its own fitted peak position,
and each distribution has unit area over its full range.}
\label{fig:dchs}
\end{figure}

Figure~\ref{fig:dchs} and Table~\ref{tab:dchs} compare the spectra and peak
widths before and after event selection. The emopt4 prediction is already
$31\%$ broader than the unselected experimental peak; after selection,
the excess reaches $74\%$.

The CC calculation is shown for two incident charge distributions. The
equilibrium fractions in silicon are approximately $93.4\%$, $6.5\%$ and
$0.1\%$ for $q=54,53,52$, respectively. A second calculation uses fractions
of $70\%$, $20\%$ and $10\%$ for the same charges to illustrate the effect
of a charge distribution altered by upstream channeling. The resulting widths,
$0.889\%$ and $0.932\%$, lie below the selected experimental width,
consistent with residual beam and detector broadening.

\begin{table}[!htb]
\caption{Relative xenon peak widths in the $300\,\mu\mathrm{m}$ DPS-NICA
pad. All calculations are at $267\,\mathrm{MeV}$ per nucleon.
Quoted uncertainties are statistical fit errors.}
\label{tab:dchs}
\begin{ruledtabular}
\begin{tabular}{lc}
Sample & $w$ (\%) \\
\hline
Experiment, no event cuts & $1.4073\pm0.0005$ \\
Experiment, selected & $1.060\pm0.008$ \\
Geant4 electromagnetic option 4 & $1.844\pm0.020$ \\
CC, equilibrium charge distribution & $0.889\pm0.010$ \\
CC, $70/20/10$ charge mixture & $0.932\pm0.010$ \\
\end{tabular}
\end{ruledtabular}
\end{table}

\subsection{STIKL signal widths}
\begin{figure}[!htb]
\centering
\includegraphics[width=\linewidth]{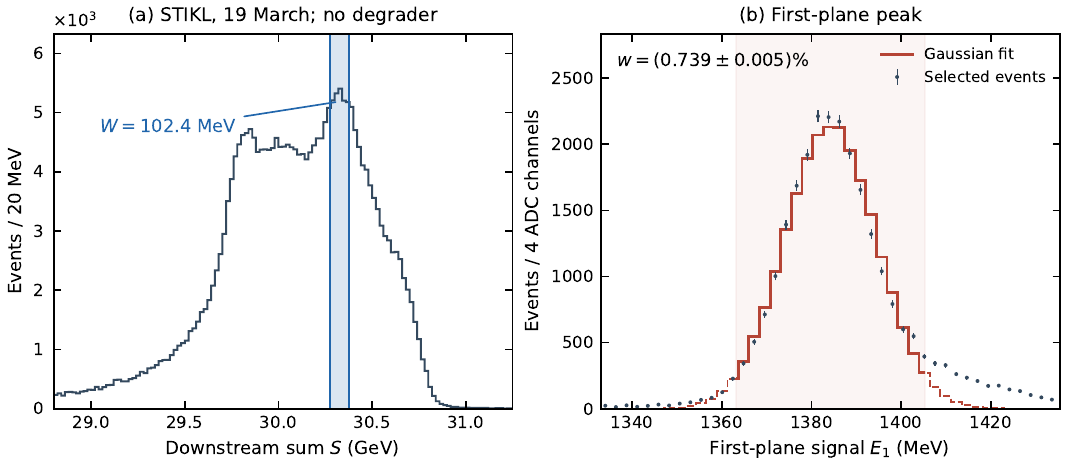}
\caption{STIKL data without a degrader, recorded on 19 March at an
incident energy of approximately $274\,\mathrm{MeV}$ per nucleon.
(a) Downstream summed-signal distribution. The shaded band is the
selection window of full width $W=102.4\,\mathrm{MeV}$ around the
higher-signal maximum. (b) First-plane signal of the retained events
and its Gaussian-core fit. For display, four adjacent ADC channels
are grouped into each bin (about $2.4\,\mathrm{MeV}$). The solid curve
shows the bin-integrated model in the fitted interval; the dashed curve
is its extrapolation.}
\label{fig:stikl_selection}
\end{figure}

Figure~\ref{fig:stikl_selection} illustrates the selection for 19 March.
The iteration gives $W=102.4\,\mathrm{MeV}$, with the centre fixed
at $S_c=30.327\,\mathrm{GeV}$. The window retains $26\,887$ events.
The fitted first-plane width is $\sigma_1=10.23\,\mathrm{MeV}$,
or $w=(0.739\pm0.005)\%$, where the uncertainty is statistical.

\begin{samepage}
The high-energy tail in Fig.~\ref{fig:stikl_selection}(b) may be associated
with channeling effects, as high-side tails have been observed in silicon
energy-deposition spectra of channeled heavy ions~\cite{Moller2001}.\par
\end{samepage}

Table~\ref{tab:stikl} lists the three selections and their fitted widths.

\begin{table}[!htb]
\caption{STIKL event selections and experimental upper estimates of the
first-plane peak width. $S_c$ is the fixed window centre and $W$ is the
full width from Eq.~\eqref{eq:stikl_gate}. The fitted standard deviation
and relative width are $\sigma_1$ and $w=\sigma_1/\mu_1$.
Incident energies are approximate. The errors quoted with $w$ are
statistical; $\delta w_{\rm sel}^{+}$ is the additional one-sided
selection systematic, expressed in percentage points.}
\label{tab:stikl}
\begin{ruledtabular}
\begin{tabular}{@{}rrrrrrr@{}}
\shortstack{$E$\\(MeV/u)} & \shortstack{$S_c$\\(MeV)} & \shortstack{$W$\\(MeV)} & Events & \shortstack{$\sigma_1$\\(MeV)} & \shortstack{$w$\\(\%)} & \shortstack{$\delta w_{\rm sel}^{+}$\\(\%)} \\
\hline
152 & 15588.1 & 143.5 & 11004 & 14.34 & $0.706\pm0.008$ & $+0.038$ \\
195 & 20868.9 & 113.9 & 17742 & 11.35 & $0.663\pm0.006$ & $+0.025$ \\
274 & 30327.0 & 102.4 & 26887 & 10.23 & $0.739\pm0.005$ & $+0.016$ \\
\end{tabular}
\end{ruledtabular}
\end{table}

\begin{samepage}
Channeling and detector nonuniformity provide further broadening.
Channeling changes silicon signal distributions~\cite{Wiedenbeck2021};
the plate thickness varies across its surface within about $1\%$.
The dead-layer thickness and its spatial variation were not measured.
Their importance for silicon-telescope resolution is illustrated by
the mapping and corrections used in CRIS~\cite{Stone1998CRIS}.
Some charge-collection loss after the substantial irradiation of STIKL
is also possible, as observed after intense xenon exposure in the BM@N
silicon beam tracker~\cite{Chemezov2026}.\par
\end{samepage}

The experimental estimates retain these external contributions, whereas
the simulations describe a monoenergetic beam at normal incidence on the
detector centre in vacuum.

\subsection{Comparison across energies}
\begin{figure}[!htb]
\centering
\includegraphics[width=0.86\linewidth]{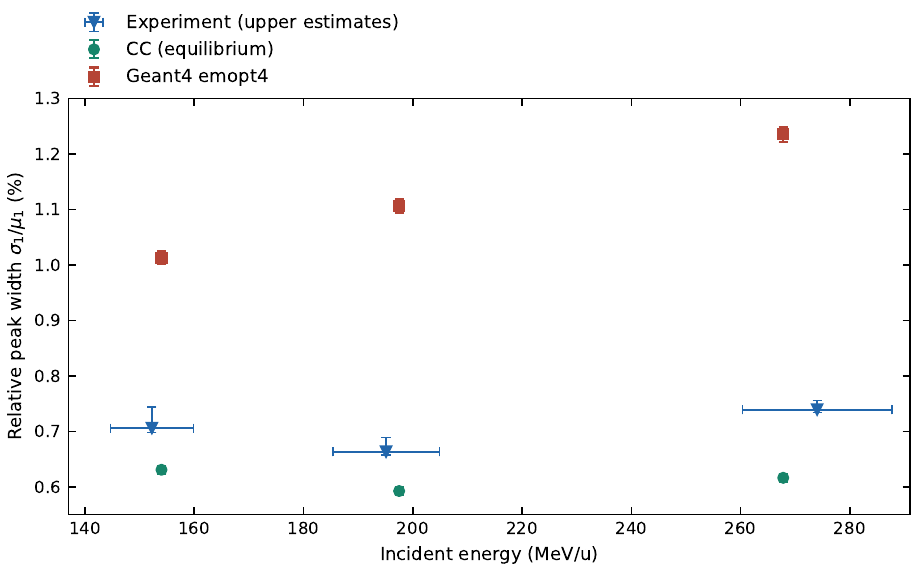}
\caption{Relative peak widths in the $675\,\mu\mathrm{m}$ first STIKL
plane. Downward triangles denote experimental upper estimates obtained
with the iterative windows of Eq.~\eqref{eq:stikl_gate}; squares and circles denote emopt4 and
CC calculations, respectively. CC samples use an equilibrium incident
charge distribution. Experimental vertical bars combine statistical
errors and the one-sided selection systematic in quadrature; the
systematic contribution extends towards larger widths. Model error
bars are statistical. Horizontal bars on the experimental points show the adopted
$\pm5\%$ systematic uncertainty in the energy scale.}
\label{fig:stikl}
\end{figure}

Figure~\ref{fig:stikl} compares the three STIKL upper estimates with
the calculations. The simulation energies of 154, 197.5 and
$267.8\,\mathrm{MeV}$ per nucleon differ from the corresponding
experimental energies by less than the adopted $5\%$ energy-scale
uncertainty. The emopt4 widths exceed the experimental estimates by
approximately $44$--$67\%$.

The CC calculations start from the equilibrium charge distribution in
silicon. Their relative widths of $0.59$--$0.63\%$ lie below the
experimental estimates throughout the studied interval, consistent
with the remaining beam and detector broadening.

\newpage
\section{Conclusions}
Xenon measurements in 300- and $675\,\mu\mathrm{m}$ silicon show that
Geant4 electromagnetic option 4 overestimates central signal widths in
the studied energy interval. In the $300\,\mu\mathrm{m}$ pad the excess
is already $31\%$ before event selection and reaches $74\%$ after selection.
In the $675\,\mu\mathrm{m}$ plane it is $44$--$67\%$ relative to
the experimental upper estimates obtained with iterative downstream-signal
windows of approximately ten first-plane standard deviations.
Calculations with explicit charge exchange give smaller widths consistent
with these upper estimates for both thicknesses. The DPS-NICA comparison
also illustrates the sensitivity of the response to the incident charge
distribution.

\section*{Funding}
The study was conducted under the state assignment of Lomonosov Moscow State University.

\section*{Conflict of interest}
The authors declare that they have no conflict of interest.


\clearpage
\section*{References}
\begin{thebibliography}{99}
\bibitem{Weick2000}
H.~Weick \textit{et al.}, ``Drastic Enhancement of Energy-Loss Straggling
of Relativistic Heavy Ions due to Charge-State Fluctuations,''
Phys. Rev. Lett. \textbf{85}, 2725--2728 (2000).
\url{https://doi.org/10.1103/PhysRevLett.85.2725}
\bibitem{Weick2002}
H.~Weick \textit{et al.}, ``Energy-loss straggling of (200--1000) MeV/u
uranium ions,'' Nucl. Instrum. Methods Phys. Res. B \textbf{193}, 1--7 (2002).
\url{https://doi.org/10.1016/S0168-583X(02)00718-8}
\bibitem{Hasebe1994}
N.~Hasebe \textit{et al.}, ``Energy Loss Straggling of Heavy Charged Particles
in Thick Silicon Absorbers,'' J. Phys. Soc. Jpn. \textbf{63}, 92--100 (1994).
\url{https://doi.org/10.1143/JPSJ.63.92}
\bibitem{Hasebe1999}
N.~Hasebe \textit{et al.},
``Straggling in Energy Loss of Energetic Heavy Ions ($Z\leq8$) in Thick
Silicon Absorber,'' J. Phys. Soc. Jpn. \textbf{68}, 1556--1561 (1999).
\url{https://doi.org/10.1143/JPSJ.68.1556}
\bibitem{Agostinelli2003}
S.~Agostinelli \textit{et al.}, Nucl. Instrum. Methods Phys. Res. A
\textbf{506}, 250 (2003).
\url{https://doi.org/10.1016/S0168-9002(03)01368-8}
\bibitem{Kurganov2024}
A.~Kurganov \textit{et al.}, ``Method of Experimental Alignment of Detectors
in the DPS-NICA Project,'' Phys. Part. Nucl. Lett. \textbf{21}, 178--185 (2024).
\url{https://doi.org/10.1134/S1547477124020080}
\bibitem{Vasiliev2021}
O.~Vasiliev \textit{et al.}, ``Current Status of the NUCLEON-2 Mission,''
Phys. Part. Nucl. Lett. \textbf{18}, 36--51 (2021).
\url{https://doi.org/10.1134/S154747712101012X}
\bibitem{ATIMA}
H.~Geissel \textit{et al.}, ATIMA, GSI software documentation.
\url{https://web-docs.gsi.de/~weick/atima/}
\bibitem{ScheidenbergerGeissel1998}
C.~Scheidenberger and H.~Geissel, Nucl. Instrum. Methods Phys. Res. B
\textbf{135}, 25--34 (1998).
\url{https://doi.org/10.1016/S0168-583X(97)00639-3}
\bibitem{WeickATIMA2017}
H.~Weick \textit{et al.}, ``Improved accuracy of the code ATIMA for
energy loss of heavy ions in matter,'' GSI Scientific Report 2017.
\url{https://web-docs.gsi.de/~weick/atima/ATIMA_GSI_scientific_report_2017-5.pdf}
\bibitem{Yang1991}
Q.~Yang, D.~J.~O'Connor, and Z.~Wang, Nucl. Instrum. Methods Phys. Res. B
\textbf{61}, 149 (1991).
\url{https://doi.org/10.1016/0168-583X(91)95454-L}
\bibitem{Valentin2012}
A.~Valentin, M.~Raine, J.-E.~Sauvestre, M.~Gaillardin, and P.~Paillet,
Nucl. Instrum. Methods Phys. Res. B \textbf{288}, 66--73 (2012).
\url{https://doi.org/10.1016/j.nimb.2012.07.028}
\bibitem{Raine2014}
M.~Raine, M.~Gaillardin, and P.~Paillet,
Nucl. Instrum. Methods Phys. Res. B \textbf{325}, 97--100 (2014).
\url{https://doi.org/10.1016/j.nimb.2014.01.014}
\bibitem{Scheidenberger1998}
C.~Scheidenberger \textit{et al.}, Nucl. Instrum. Methods Phys. Res. B
\textbf{142}, 441--462 (1998).
\url{https://doi.org/10.1016/S0168-583X(98)00244-4}
\bibitem{Lijian1995}
T.~Lijian, H.~Qing, and L.~Zhengming, Radiat. Phys. Chem.
\textbf{45}, 235--245 (1995).
\url{https://doi.org/10.1016/0969-806X(94)00063-8}
\bibitem{CompanionPackage}
A.~Azhakin and I.~Kudryashov, ``CCGeant4: charge-state-resolved transport
and energy-loss fluctuations of relativistic heavy ions in Geant4,''
arXiv:2611.15759 (2026).
\url{https://arxiv.org/abs/2611.15759}
\bibitem{Moller2001}
S.~Pape M{\o}ller \textit{et al.}, ``Random and channeled energy loss
of 33.2-TeV Pb nuclei in silicon single crystals,''
Phys. Rev. A \textbf{64}, 032902 (2001).
\url{https://doi.org/10.1103/PhysRevA.64.032902}
\bibitem{Wiedenbeck2021}
M.~E.~Wiedenbeck \textit{et al.}, ``Thin silicon solid-state detectors
for energetic particle measurements: Development, characterization,
and application on NASA's Parker Solar Probe mission,''
Astron. Astrophys. \textbf{650}, A27 (2021).
\url{https://doi.org/10.1051/0004-6361/202039754}
\bibitem{Stone1998CRIS}
E.~C.~Stone \textit{et al.}, ``The Cosmic-Ray Isotope Spectrometer for
the Advanced Composition Explorer,'' Space Sci. Rev. \textbf{86},
285--356 (1998).
\url{https://doi.org/10.1023/A:1005075813033}
\bibitem{Chemezov2026}
D.~Chemezov \textit{et al.}, ``Performance and Radiation Damage of the
Heavy-Ion Silicon Beam Tracker in the BM@N Experiment,''
arXiv:2610.17947 (2026).
\url{https://doi.org/10.48550/arXiv.2610.17947}
\end{thebibliography}
\end{document}